\documentclass[conference]{IEEEtran}
\IEEEoverridecommandlockouts
\usepackage{cite}
\usepackage{amsmath,amssymb,amsfonts}
\usepackage{algorithmic}
\usepackage{graphicx}
\usepackage{textcomp}
\usepackage{xcolor}
\usepackage{booktabs}
\usepackage{multirow}
\usepackage{hyperref}
\usepackage{listings}
\usepackage[T1]{fontenc}
\usepackage{inconsolata}
\usepackage{verbatim}
\usepackage{caption}
\usepackage{subcaption}
\usepackage{enumitem}

\def\BibTeX{{\rm B\kern-.05em{\sc i\kern-.025em b}\kern-.08em
    T\kern-.1667em\lower.7ex\hbox{E}\kern-.125emX}}

\definecolor{belizehole}{HTML}{2980b9}
\definecolor{clouds}{HTML}{f4f8f9}
\definecolor{midnightblue}{HTML}{2c3e50}
\definecolor{pomegranate}{HTML}{c0392b}
\definecolor{royalblue}{HTML}{3867d6}
\definecolor{greensea}{HTML}{16a085}
\definecolor{nephritis}{HTML}{27ae60}
\definecolor{amethyst}{HTML}{9b59b6}

\lstdefinestyle{interfaces}{
	float=t,
	floatplacement=t,
}

\begin{document}

\title{Improving Scalability with GPU-Aware Asynchronous Tasks}

\author{\IEEEauthorblockN{Jaemin Choi\IEEEauthorrefmark{1}, David F.~Richards\IEEEauthorrefmark{2}, Laxmikant V.~Kale\IEEEauthorrefmark{1}}
\IEEEauthorblockA{\IEEEauthorrefmark{1}Department of Computer Science, University of Illinois at Urbana-Champaign, Urbana, Illinois, USA}
\IEEEauthorblockA{\IEEEauthorrefmark{2}Center for Applied Scientific Computing, Lawrence Livermore National Laboratory, Livermore, California, USA\\
	Email: \{jchoi157,kale\}@illinois.edu,
	richards12@llnl.gov}}


\maketitle

\begin{abstract}
Asynchronous tasks, when created with over-decomposition, enable automatic computation-communication overlap which can substantially improve performance and scalability. This is not only applicable to traditional CPU-based systems, but also to modern GPU-accelerated platforms. While the ability to hide communication behind computation can be highly effective in weak scaling scenarios, performance begins to suffer with smaller problem sizes or in strong scaling due to fine-grained overheads and reduced room for overlap. In this work, we integrate GPU-aware communication into asynchronous tasks in addition to computation-communication overlap, with the goal of reducing time spent in communication and further increasing GPU utilization. We demonstrate the performance impact of our approach using a proxy application that performs the Jacobi iterative method, Jacobi3D. In addition to optimizations to minimize synchronizations between the host and GPU devices and increase the concurrency of GPU operations, we explore techniques such as kernel fusion and CUDA Graphs to mitigate fine-grained overheads at scale.
\end{abstract}


\section{Introduction}


GPUs are driving today's research in many key areas including computational science, machine learning, data analytics, and cloud computing.
In molecular biology, Team \#COVIDisAirborne has utilized 4,096 nodes (24,576 GPUs) of the Summit supercomputer to perform a data-driven simulation of the SARS-CoV-2 Delta variant, providing unprecedented atomic-level views of the virus in a respiratory aerosol~\cite{sc21-covid}.
In addition to the myriad of GPU-accelerated systems in the current TOP500 list of the world's fastest supercomputers~\cite{top500}, upcoming U.S. Department of Energy exascale systems such as Aurora~\cite{aurora} and Frontier~\cite{frontier} will rely on next-generation GPUs for the bulk of their computing horsepower. In industry, Meta has recently announced the AI Research SuperCluster (RSC), which will be used to train large models in natural language processing (NLP) and computer vision, paving the way for the metaverse~\cite{meta-rsc}. Meta plans to increase the number of NVIDIA A100 GPUs in RSC from 6,080 to 16,000, which will provide nearly five exaFLOPS of mixed precision compute.

The sheer degree of computational power and data parallelism provided by GPUs are enabling applications to achieve groundbreaking performance. However, due to the relatively slower improvement of network bandwidth compared to the computational capabilities of GPUs over time, communication overheads often hold applications back from achieving high compute utilization and scalability. Overlapping computation and communication is a widely used technique to mitigate this issue, but it is generally up to the application programmer to identify potential regions of overlap and implement the necessary mechanisms. This becomes increasingly difficult in applications with convoluted code structures and interleavings of computation and communication.
\textit{Automatic} computation-communication overlap can be achieved with overdecomposition and asynchronous task execution, features supported by the Charm++ runtime system and its family of parallel programming models~\cite{charm},
substantially improving performance and scalability on both CPU and GPU based systems~\cite{espm220-choi}.

However, performance gains from overdecomposition-driven overlap can degrade with finer task granularity. In weak scaling scenarios with a small base problem size or at the limits of strong scaling, fine-grained overheads associated with communication, scheduling, and management of GPU operations can outweigh the benefits from computation-communication overlap.
In this work, we propose the integration of GPU-aware communication into asynchronous execution of overdecomposed tasks, to reduce communication overheads and enable higher degrees of overdecomposition at scale.
In addition to improving performance and scalability, overdecomposition enables adaptive runtime features such as load balancing and fault tolerance. Asynchronous execution of overdecomposed tasks also provide the benefit of spreading out communication over time, allowing more efficient use of the network when bandwidth is limited~\cite{thesis20-robson}.

We also demonstrate the importance of minimizing synchronizations between the host and device and increasing the concurrency of independent GPU operations, by comparing the performance of a proxy application against the implementation described in our previous work~\cite{espm220-choi}.
In addition to these optimizations, we explore techniques such as kernel fusion~\cite{greencom10-kernel_fusion} and CUDA Graphs~\cite{cuda_graphs} to mitigate overheads related to fine-grained GPU execution, which can be exposed at the limits of strong scaling. We show how these mechanisms improve performance especially for relatively high degrees of overdecomposition, which can be useful for taking advantage of runtime adaptivity.

The major contributions of this work can be summarized as the following:

\begin{itemize}
	\item We present the integration of overdecomposed asynchronous tasks and GPU-aware communication to exploit computation-communication overlap and reduce exposed communication overheads.
	\item We demonstrate the impact of our approach by evaluating the weak and strong scaling performance of a scientific proxy application on a large-scale GPU-accelerated system.
	\item We illustrate the importance of minimizing synchronizations between the host and device as well as ensuring concurrency of independent GPU operations.
	\item We explore kernel fusion and CUDA Graphs as techniques to reduce fine-grained overheads at scale and evaluate their impact on performance.
\end{itemize}


\section{Background}\label{sec:background}

\subsection{Automatic Computation-Communication Overlap}

Overlapping computation and communication is a widely used and researched technique,
which has been proven to be effective in both CPU-based and GPU-accelerated systems
for hiding communication latency~\cite{espm220-choi}.
Non-blocking communication is one of the primary mechanisms used to
expose opportunities for overlap, allowing processors to perform useful work while
communication is being progressed~\cite{sc07-nonblocking_mpi}.
With the Message Passing Interface (MPI), a distributed memory communication standard broadly used in HPC,
it is the application programmer's responsibility
to identify regions of potential overlap~\cite{paraplop09}. Not only is this often
challenging due to complex code structure and flow of execution, but it also limits
the amount of attainable overlap to the identified regions.

\begin{figure}[t]
\centering
\begin{subfigure}{\linewidth}
\centering
\includegraphics[width=0.4\linewidth]{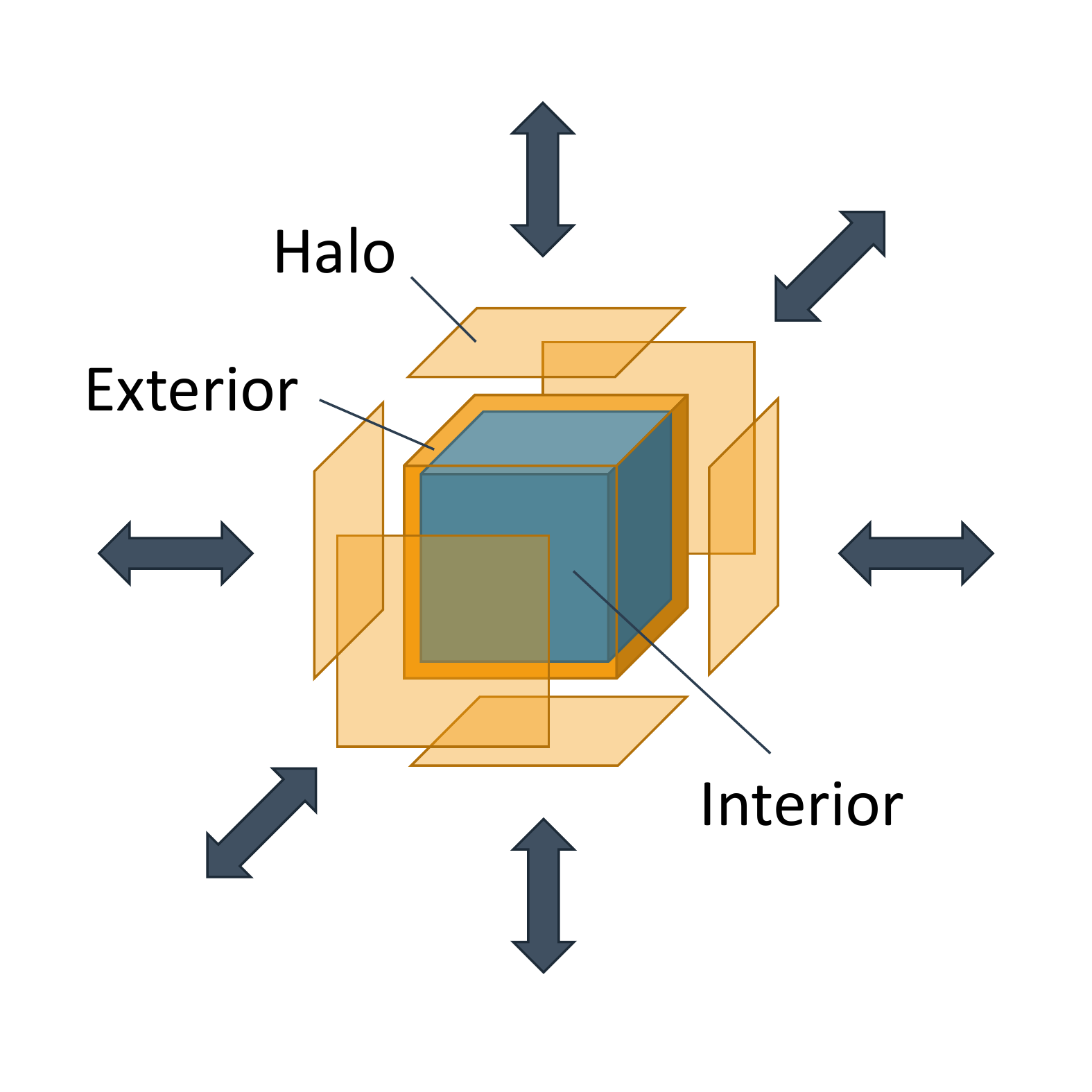}
\caption{Diagram}
\label{fig:mpi_example_diagram}
\end{subfigure}
\hfill
\vspace{10pt}
\begin{subfigure}{\linewidth}
\begin{lstlisting}[language=C++]
for (int iter = 0; iter < n_iters; iter++) {
  // Pack halo data for sending
  packHalos(send_halo, ...);

  // Post non-blocking receives and sends
  for (int dir = 0; dir < 6; dir++) {
      MPI_Irecv(recv_halo[dir], ..., requests[2*dir]);
      MPI_Isend(send_halo[dir], ..., requests[2*dir+1]);
  }

  if (overlap) {
    // Perform Jacobi update on the interior
    interiorUpdate();
  }

  // Wait for halo exchanges to complete
  MPI_Waitall(12, requests, statuses);

  // Unpack received halos to the 3D block
  unpackHalos(recv_halo, ...);

  if (overlap) {
    // Perform Jacobi update on the exterior
    exteriorUpdate();
  } else {
    // Perform Jacobi update on the whole block
    update();
  }
}
\end{lstlisting}
\caption{Code}
\label{fig:mpi_example_code}
\end{subfigure}
\caption{MPI 3D Jacobi example (Jacobi3D) with a manual overlap option.
The non-blocking MPI
communication can overlap with the interior Jacobi update which is
independent of the halo data coming from the neighbors.}
\label{fig:mpi_example}
\vspace{-10pt}
\end{figure}

For example, let us have a look at how
a three-dimensional Jacobi iterative method, hereafter called Jacobi3D, can be implemented using MPI.
Each MPI process is responsible for a block of the global 3D grid, as described in 
Figure~\ref{fig:mpi_example_diagram}.
Halo data are first exchanged among the neighbors using non-blocking MPI
communication routines. After all halo data are received and unpacked, each MPI process can perform
the Jacobi update on its block.
However, since updating only the interior of the block
does not depend on the neighbors' halo data, it can overlap with the halo exchanges.
Implementations with and without this manual overlap are described in Figure~\ref{fig:mpi_example_code}.
Finding such regions of potential overlap, however, can be much more challenging in larger applications.
Furthermore, the execution could be blocked at synchronization points (e.g., ~\texttt{MPI\_Waitall})
if such calls are made too early, limiting the amount of attainable overlap.
Periodically polling for the completion of the communication routines is an alternative,
but it is not compatible with the sequential execution flow of typical MPI applications
and can also unnecessary consume CPU cycles~\cite{ics19-overlap}.

\begin{figure}[t]
\centering
\includegraphics[width=0.85\linewidth]{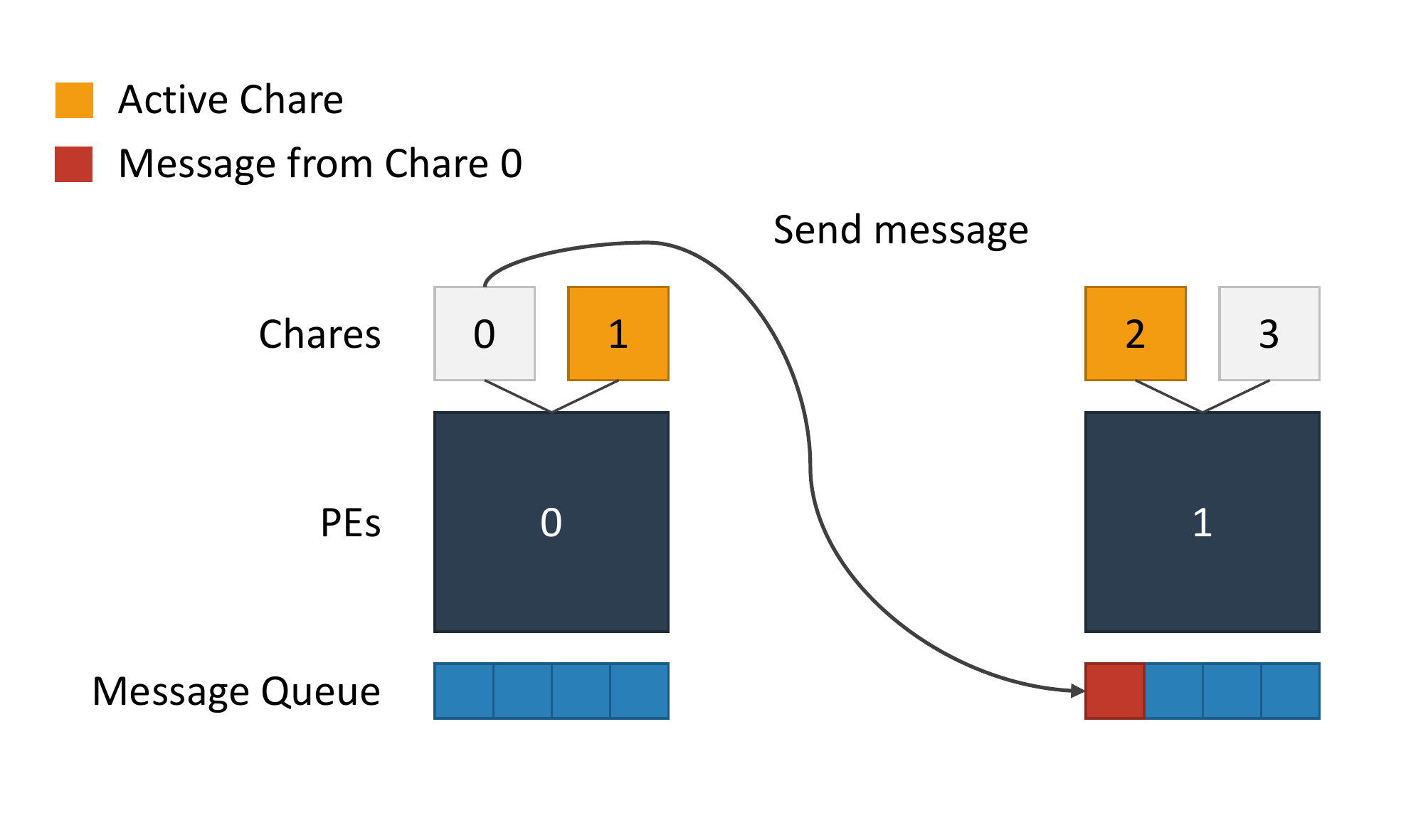}
\caption{Message-driven execution model in Charm++.
\mbox{Chare 0} sending a message to \mbox{Chare 2} can overlap with computation of \mbox{Chare 1}.}
\label{fig:charm_exec_model}
\vspace{-10pt}
\end{figure}

\textit{Automatic} computation-communication overlap relieves the programmer from
the responsibility of manually orchestrating tasks for overlap. It is achieved in
the Charm++ parallel programming system~\cite{charm} on the foundation of two core features:
\textbf{overdecomposition} and \textbf{asynchronous task execution}. In a Charm++ program, the problem domain
can be decomposed into more units of work and/or data, called \textit{chares}, than the number of available
processing elements (PEs). This is in contrast to conventional MPI applications where
a single MPI process is assigned to each PE. In addition to being able to automatically
overlap computation of one chare object with communication of another\footnote{Computation and communication of the same chare can also be overlapped, as long as they are asynchronous.}, overdecomposition
empowers the runtime system to support adaptive features such as dynamic load balancing
and fault tolerance. Another benefit of overdecomposition is that the injection of messages
into the network can be spread out over time, alleviating pressure on the network~\cite{thesis20-robson}.

As shown in Figure~\ref{fig:charm_exec_model}, Charm++ employs an asynchronous message-driven
execution model where the arrival of a message triggers a certain task of the target chare to
be executed.
This message encapsulates information about which C++ method of
the target chare, i.e., \textit{entry method}, should be executed, along with
the necessary data.
Incoming messages are accumulated in a message queue that is continuously
checked by a scheduler that runs on each PE.
The execution of a Charm++ application begins with
the Main Chare, which is defined by the user to play the role similar to that of the main function in regular C++.
The Main Chare can can create other chare objects and initiate the flow of execution by invoking their entry methods.
The invocation of a chare's entry method translates into a message transmission by the runtime system,
which is by default asynchronous. This increases opportunities for computation-communication overlap
by allowing the execution to continue after minimal processing when a Charm++ communication primitive is called.
Once a chare entry method finishes executing, the scheduler will pick up another message from the queue to execute the next entry method.

Reducing unnecessary synchronization, between work units (chares in Charm++)
as well as between the host and GPU devices,
is another critical factor in exploiting computation-communication overlap.
Asynchronous execution can minimize idle time and expose more opportunities
for overlap by allowing each work unit to progress as freely as possible,
enforcing only the necessary dependencies between tasks.
Taking Jacobi3D as an example,
it is not necessary to perform a global synchronization across all work units
after every iteration; in fact, each unit only needs to ensure that it is exchanging
halo data from the same iteration with its neighbors.
On NVIDIA GPUs, kernel launches and data transfers can be made asynchronous with the use of CUDA
Streams, allowing work to be offloaded to the GPU without blocking the progress
of the host CPU. However, asynchronously detecting the completion of GPU tasks
requires a different mechanism especially for scheduler-driven runtime systems
such as Charm++, which is discussed in Section~\ref{sec:design_overlap}.

\begin{figure}[t]
\centering
\begin{subfigure}{\linewidth}
\begin{lstlisting}[language=C++, morekeywords={readonly,mainchare,entry,array,serial,when}]
readonly CProxy_Block block_proxy;

mainchare Main {
  entry Main(...);
};

array [3D] Block {
  entry void recvHalo(...);
  entry void run() {
  	for (iter = 0; iter < n_iters; iter++) {
      // Pack and send halo data to neighbors
      serial { packHalos(); sendHalos(); }

      // Asynchronously receive halo data and unpack
      for (count = 0; count < 6; count++) {
        when recvHalo[iter](int r, double* h, int s) {
          serial { unpackHalo(); }
        }
      }

      // Perform Jacobi update on the entire block
      serial { update(); }
    }
  }
};
\end{lstlisting}
\caption{.ci file}
\label{fig:charm_example_ci}
\end{subfigure}
\hfill
\vspace{10pt}
\begin{subfigure}{\linewidth}
\begin{lstlisting}[language=C++]
void Main::Main(...) {
  // Create a 3D indexed array of chares
  block_proxy = ckNew(n_x, n_y, n_z, ...);

  // Start the simulation by invoking an entry
  // method on all block chares (broadcast)
  block_proxy.run();
}

void BlockChare::sendHalos() {
  // Send halos to neighbors by using proxy
  for (int dir = 0; dir < 6; dir++) {
    block_proxy(idx[dir]).recvHalo(iter, halo[dir], size);
  }
}

void BlockChare::packHalos() { ... }
void BlockChare::unpackHalo() { ... }
void BlockChare::update() { ... }
\end{lstlisting}
\caption{.C file}
\label{fig:charm_example_c}
\end{subfigure}
\caption{Charm++ version of Jacobi3D with automatic overlap. The Charm
Interface (CI) file contains user-declared components that relate
to parallel execution, including chares, entry methods, and proxies.}
\label{fig:charm_example}
\vspace{-10pt}
\end{figure}

Figure~\ref{fig:charm_example} describes the code for a Charm++ version
of Jacobi3D. The Charm Interface (CI) file in Figure~\ref{fig:charm_example_ci}
is written by the user to declare components of parallel execution
such as chares, entry methods, and proxies.
Other codes including function bodies can be written in regular C++.
The execution begins with \texttt{Main::Main} on \mbox{PE 0},
where an indexed collection of chares, called a \textit{chare array}, is created.
By default, Chares are distributed to all the available PEs using a block mapping;
if a chare array of size eight is created on two PEs,
each PE will be responsible for four consecutive chare elements.
The creation of chares returns a handle to their proxy, which is used for
invoking entry methods.
For example,
calling \texttt{block\_proxy(0,0,0).run} will invoke the \texttt{run}
entry method on that element of the 3D chare array.
An entry method invocation on the entire proxy (e.g., \texttt{block\_proxy.run})
will perform a broadcast to invoke the same entry method on all chare elements managed by that proxy.

In Charm++ Jacobi3D, the overall flow of parallel execution
is encapsulated in the \texttt{Block::run} entry method.
Its body is composed using Structured Dagger (SDAG)~\cite{sdag}, which prevents
the program sequence from becoming obscured by the message-driven nature
of Charm++.
The \texttt{serial} construct wraps regular C++ code including function calls,
and the \texttt{when} construct allows the calling chare
to asynchronously wait for message arrivals.
Reference numbers are used in Jacobi3D to match the iteration number of an incoming message (\texttt{r} in \texttt{recvHalo})
with the block's (\texttt{iter}), to ensure that blocks progress in step with its neighbors.
Control is returned back to the scheduler at the execution of the \texttt{when}
construct, allowing other messages to be processed.
Once an awaited message arrives, the runtime system schedules the
designated entry method (e.g., \texttt{recvHalo}) to be executed.

\subsection{GPU-Aware Communication}\label{sec:back_gpu_comm}

Without support for GPU memory from the underlying communication library,
applications need explicit host-device data transfers to stage GPU buffers on host memory for communication.
Not only do such host-staging methods require more code, but they also suffer from longer latency
and reduction in attainable bandwidth.
GPU-aware communication aims to mitigate these issues,
addressing both programmer productivity and communication performance.

CUDA-aware MPI implements GPU-aware communication for NVIDIA GPUs in MPI, by supporting
GPU buffers as inputs to its communication API.
This not only eases programming by obviating the need for explicit
host-device data transfers, but also improves performance
by directly moving data between the GPU and Network Interface Card (NIC).
GPUDirect~\cite{gpudirect, gpudirect_rdma} is one of the core technologies
that drive GPU-aware communication, providing direct GPU memory access to the NIC.

In Charm++, there are two available mechanisms for GPU-aware communication: GPU Messaging API
and Channel API. The GPU Messaging API retains the message driven execution model but
requires an additional metadata message to arrive before the receiver is able to post
the receive for the incoming GPU buffer. The metadata message also invokes a \textit{post entry method}
on the receiver, which is used to inform the runtime system where the destination GPU buffer is located~\cite{ashes21-choi}.
The Channel API has been recently developed to address the performance issues with this mechanism,
which uses two-sided send and receive semantics for efficient data movement~\cite{arxiv22-choi}.
It should be noted that both APIs use the Unified Communication X (UCX) library~\cite{ucx_paper} as a low-level interface.
In this work, the Channel API is used to drive GPU-aware communication in Charm++,
with its implementation in Jacobi3D discussed in Section~\ref{sec:design_gpu_comm}.

\section{Design and Implementation}\label{sec:design}

We propose the integration of GPU-aware communication in
asynchronous tasks created with overdecomposition
to improve application performance and scalability.
In addition to a detailed discussion on combining these two mechanisms,
we describe optimizations to the baseline Jacobi3D proxy application
for reducing synchronization and improving concurrency of GPU operations.
Furthermore, we explore techniques for fine-grained GPU tasks such as kernel fusion
and CUDA Graphs to mitigate potential performance issues with strong scaling.
It should be noted that although this work uses terminology from NVIDIA GPUs
and CUDA, most discussions also apply to GPUs from other vendors.

\subsection{Achieving Automatic Overlap on GPU Systems}\label{sec:design_overlap}

\begin{figure}[t]
\centering
\includegraphics[width=0.85\linewidth]{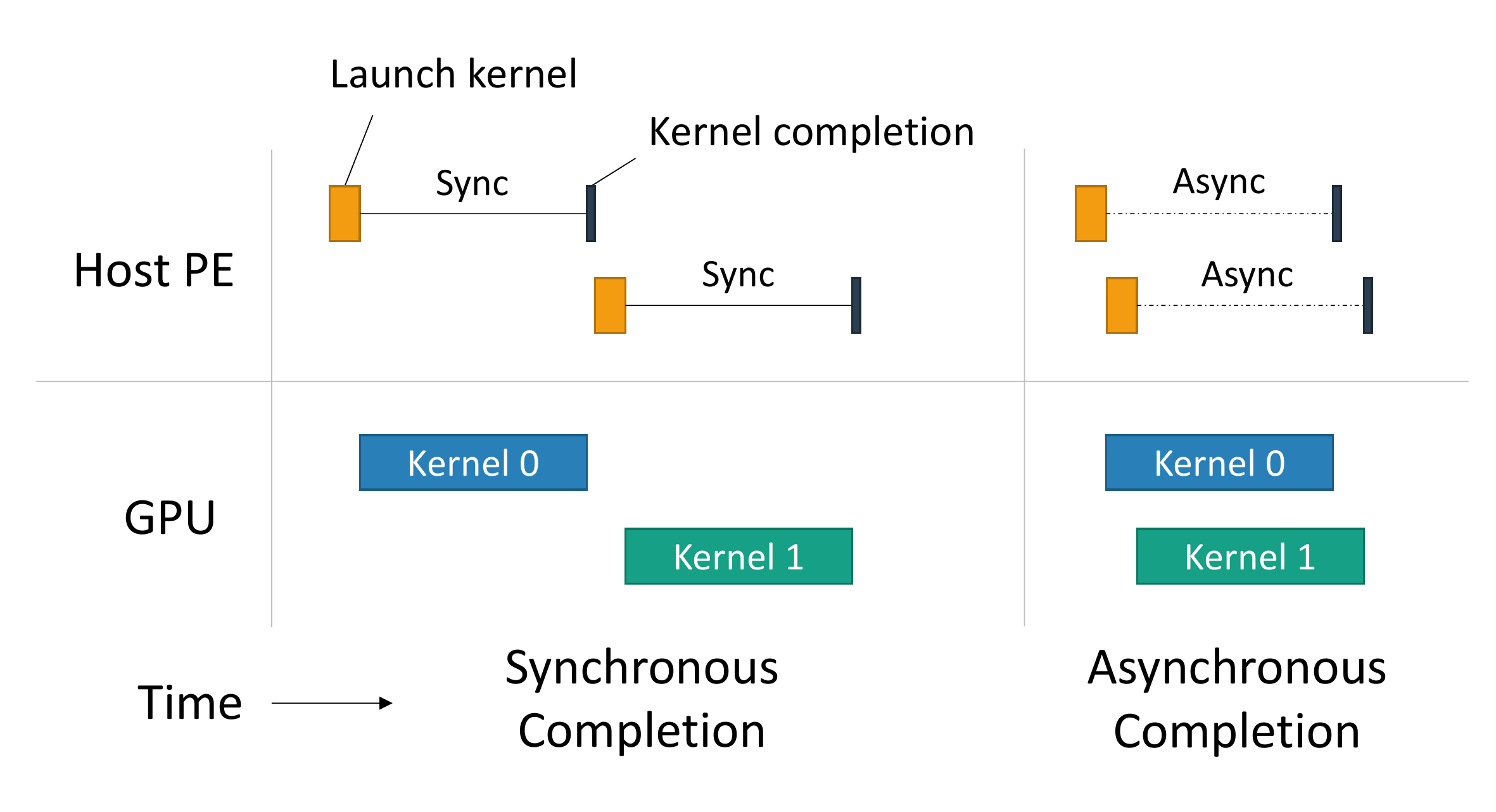}
\caption{Execution timelines demonstrating the benefits of asynchronous completion in Charm++.
This assumes that the two GPU kernels are small enough to execute concurrently on the same GPU.}
\label{fig:hapi_comparison}
\vspace{-10pt}
\end{figure}

We use Charm++ as the vehicle to achieve automatic
computation-communication overlap in GPU-accelerated execution.
Allowing GPU work to progress asynchronously and detecting
their completion as early as possible are equally important
in creating opportunities for overlap.
CUDA Streams~\cite{cuda_streams}, which allows GPU operations to execute
asynchronously and concurrently, is the preferred method of offloading
work to GPUs in Charm++ applications.
A common usage of a CUDA stream involves
enqueueing GPU work such as a kernel or memcpy and waiting for it
to finish using a synchronization mechanism, e.g., \texttt{cudaStreamSynchronize}.
Since submitting work to a CUDA stream is asynchronous, other tasks can be
performed on the host CPU until the synchronization point.
While this may be sufficient for traditional MPI applications where a single process
runs on each PE, it can be detrimental to scheduler-driven tasking frameworks
such as Charm++; synchronization can prevent the scheduler from
processing other available messages and performing useful work.
Figure~\ref{fig:hapi_comparison} compares the execution timelines
with synchronous and asynchronous completion mechanisms in Charm++, respectively,
where two chares mapped to a single PE are offloading work to the same GPU.
Asynchronous completion frees up the host CPU to perform
other tasks while GPU work is being executed, facilitating overlap.

Hybrid API (HAPI)~\cite{hapi} enables asynchronous completion
detection of GPU operations in Charm++, using CUDA events to track
their status in the scheduler. It allows the user to specify
which Charm++ method should be executed when the completion of
the tracked GPU work is detected. Meanwhile, the
scheduler can perform other useful tasks, increasing opportunities for
computation-communication overlap.
More implementation details of HAPI can be found in
our previous work~\cite{espm220-choi}.
In the optimized version of Jacobi3D as described in Section~\ref{sec:design_opt},
HAPI is used to ensure that the Jacobi update and packing kernels
have been completed before sending halo data to the neighbors.

In addition to asynchronous completion detection, prioritizing
communication and related GPU operations (e.g., packing and unpacking
kernels) is key to exploiting overlap.
Since multiple chares can utilize the same GPU
concurrently due to overdecomposition, communication-related operations of one chare can be
impeded by computational kernels launched by other chares unless they are given
higher priority. Such delays in communication translate directly
into performance degradation~\cite{espm220-choi}.
In Jacobi3D, host-device transfers and (un)packing
kernels are enqueued into high-priority CUDA streams.
The Jacobi update kernel utilizes a separate stream with
lower priority.
These streams are created for every chare object so that independent tasks from
different chares can execute concurrently on the GPU when possible.

\subsection{GPU-Aware Communication in Charm++}\label{sec:design_gpu_comm}

Exploiting computation-communication overlap with overdecomposition can be
highly effective in weak scaling scenarios where performance improvements from overlap
outweigh the overheads from finer-grained tasks. With small problem sizes
or with strong scaling, however, overdecomposition can quickly reach its limits
as task granularity decreases.
One of the main sources of overhead with fine-grained tasks is
communication, as the ratio of computation to communication diminishes and
subsequently less communication can be hidden behind computation.
GPU-aware communication can mitigate such overheads by utilizing
the networking hardware more efficiently.

\begin{figure}[t]
\centering
\begin{lstlisting}[language=C++, morekeywords={when,serial}]
/* C file */
// Create Charm++ callback to be invoked when
// a channel send or recv completes
CkCallback cb = CkCallback(CkIndex_Block::callback(), ...);

// Non-blocking sends and receives of halo data
for (int dir = 0; dir < 6; dir++) {
  channels[dir].send(send_halo[dir], size, cb);
  channels[dir].recv(recv_halo[dir], size, cb);
}

/* .ci file */
// When a Charm++ callback is invoked, check if it means
// completion of a receive and unpack if so
for (count = 0; count < 12; count++) {
  when callback() serial { if (recv) processHalo(); }
}
\end{lstlisting}
\caption{Usage of Channel API in Charm++ Jacobi3D.}
\label{fig:channel_example}
\vspace{-10pt}
\end{figure}

As described in Section~\ref{sec:back_gpu_comm}, Charm++ offers two mechanisms
for GPU-aware communication: GPU Messaging API and Channel API.
As the communication pattern in Jacobi3D is regular,
the Channel API can be easily used to exchange halo data with two-sided sends and receives.
Figure~\ref{fig:channel_example} demonstrates the usage of the Channel API in Jacobi3D,
where a communication channel is established between each pair of neighboring chares.
Send and receive calls are made to the channel to transfer halo buffers on the
GPU, which are translated into calls to the underlying UCX library.
A Charm++ callback is passed to the channel primitives to invoke an entry method
upon completion, enabling asynchronous completion detection and facilitating
computation-communication overlap.

\subsection{Optimizations to Baseline Performance}\label{sec:design_opt}

\begin{figure}[t]
\centering
\begin{subfigure}{\linewidth}
\centering
\includegraphics[width=0.8\linewidth]{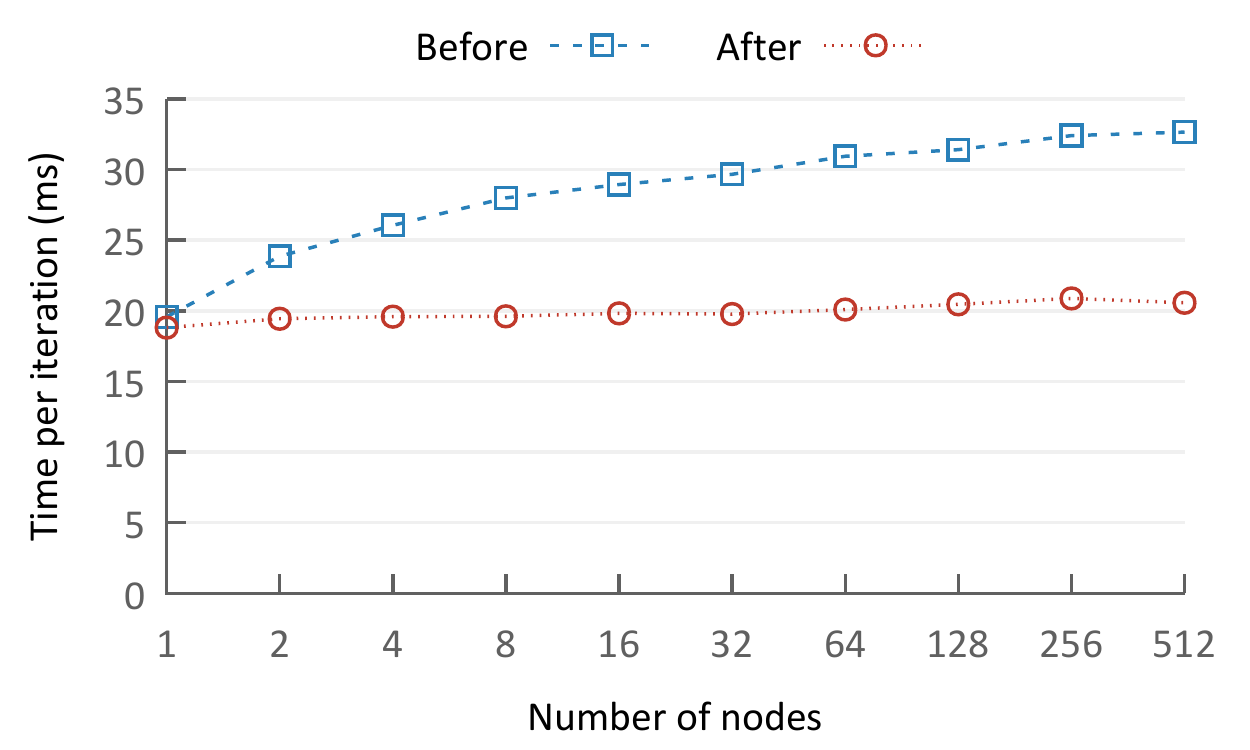}
\caption{Weak scaling, with block size of 1536 $\times$ 1536 $\times$ 1536 per node}
\label{fig:jacobi3d_baseline_weak}
\end{subfigure}
\hfill
\vspace{10pt}
\begin{subfigure}{\linewidth}
\centering
\includegraphics[width=0.8\linewidth]{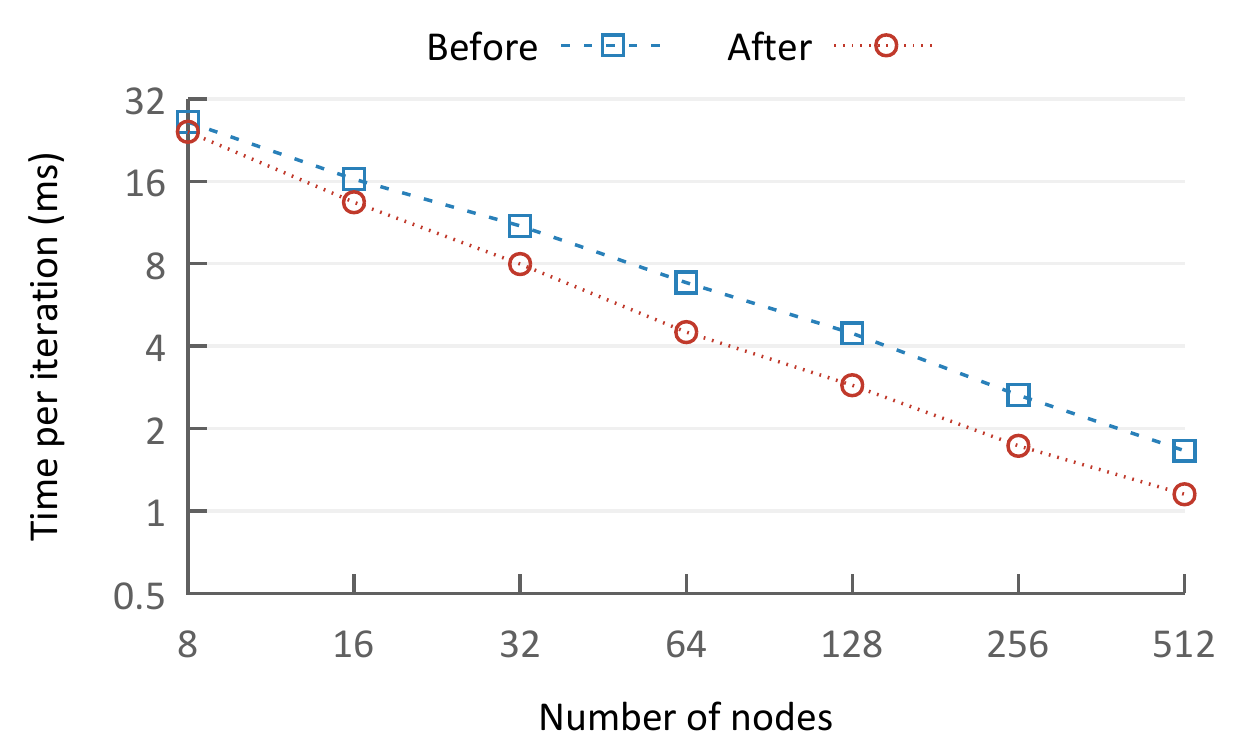}
\caption{Strong scaling, with global grid size of 3072 $\times$ 3072 $\times$ 3072 (same global grid size as on 8 nodes with weak scaling)}
\label{fig:jacobi3d_baseline_strong}
\end{subfigure}
\caption{Performance comparison of Charm++ Jacobi3D with host-staging communication before and after optimizations
on the Summit supercomputer.}
\label{fig:jacobi3d_baseline}
\vspace{-10pt}
\end{figure}


The original implementation of Jacobi3D~\cite{espm220-choi} performed a host-device synchronization
right after launching the Jacobi update kernel, to ensure that the update
is complete before incrementing the iteration counter and swapping the pointers
to the GPU buffers. Note that Jacobi3D maintains two separate buffers in GPU memory
to be used as input and output for the Jacobi update kernel.
However, this synchronization step is redundant, as the above operations to prepare for the next iteration
can instead be performed just before the halo exchanges.
This optimization reduces the number of host-device synchronizations per iteration
from two (after Jacobi update and before halo exchanges) to one (before halo exchanges).

By profiling the performance of Jacobi3D with NVIDIA Nsight Systems,
we observe that there is another optimization opportunity to increase
the concurrency of independent GPU operations. Instead of enqueueing device-host transfers
and (un)packing kernels to the same stream, we create two additional high-priority
streams for data transfers, one for device-to-host and another for host-to-device.
This allows (un)packing kernels to overlap with the data transfers, as well as the bi-directional
transfers to overlap with one another.
Unfortunately, this optimization makes enforcing dependencies between the streams more complicated.
Figure~\ref{fig:jacobi3d_baseline} showcases the improvements from the above optimizations in weak and strong scaling
performance of Charm++ Jacobi3D, with host-staging communication and a four-times overdecomposition.
All the following experiments use this new baseline implementation for various MPI and Charm++
versions of Jacobi3D.

\subsection{Techniques for Fine-grained GPU Tasks}

Strong scaling increases the amount of computational resources, e.g., number of GPUs, while maintaining the same problem size.
Consequently, the size of work and data assigned to each resource decreases as the problem is scaled out.
In GPU-accelerated environments, this causes the proportion of kernel launch overheads in execution time to grow.
Applying overdecomposition, either for computation-communication overlap or runtime adaptivity (e.g., load balancing), can exacerbate this issue.
We explore techniques such as kernel fusion~\cite{greencom10-kernel_fusion} and CUDA Graphs~\cite{cuda_graphs} to mitigate this problem in the context of fine-grained GPU execution.

Kernel fusion combines multiple kernels as a single kernel to reduce the aggregate kernel launch latency.
CUDA Graphs is a mechanism for NVIDIA GPUs where an executable graph can be constructed from multiple consecutive GPU operations, including kernels and memory copies, to reduce launch overheads. It can also expose opportunities for optimization as all necessary dependencies are presented to the CUDA runtime.
These two techniques can be used together; kernel fusion can be applied to reduce the total number of kernels, and CUDA Graphs can capture all such kernel launches and other GPU operations for more efficient repeated execution of the same graph.

\subsubsection{Kernel Fusion}\label{sec:design_kernel_fusion}

With Jacobi3D, we explore three different strategies for kernel fusion, with the fused kernels outlined below:
\begin{enumerate}[label=(\Alph*)]
	\item Packing kernels
	\item Packing kernels and unpacking kernels (as two separate kernels)
	\item Unpacking kernels, Jacobi update kernel, and packing kernels (all as a single kernel)
\end{enumerate}

Note that packing kernels can be launched right after the Jacobi update kernel, but each unpacking kernel can only be launched after the corresponding halo data arrives from a neighbor. Thus the fused version of the unpacking kernels can only be launched after all halo data arrive. When fusing the packing/unpacking kernels, the total number of GPU threads is computed as the \textit{maximum} of the different halo sizes. Each thread consecutively looks at the six faces that could be copied out as halo data, and if its index is smaller than the halo size, performs a copy into the respective halo buffer. We have found this implementation to be faster than having the total number of GPU threads to be the \textit{sum} of the halo sizes, which allows all faces to be processed concurrently but suffers from excessive control divergence. Fusing all kernels using Strategy C effectively results in one kernel execution per iteration, a significant reduction in the number of kernel launches. In this work, kernel fusion is only used in concert with GPU-aware communication to avoid complications with host-device transfers and their ensuing dependencies.

\subsubsection{CUDA Graphs}


We build a CUDA graph in Jacobi3D by capturing the entire flow of kernel launches at initialization time. The graph contains all dependencies and potential concurrency of unpacking kernels, Jacobi update kernel, and packing kernels; this simplifies each iteration of Jacobi3D to be the halo exchange phase followed by the launch of a CUDA graph.
An issue that we encountered when implementing CUDA Graphs in Jacobi3D is the limitation that parameters passed to the GPU operations in a CUDA graph should not change during execution. This is problematic since the two pointers referring to input and output data need to be swapped every iteration. Although nodes in a CUDA graph can be individually updated to use a different parameter, this is infeasible in Jacobi3D since the graph needs to be updated every iteration, nullifying the performance benefits. Our solution was to create two separate CUDA graphs, one with the two pointers reversed to the other, and alternate between them for each iteration.
As with kernel fusion, CUDA Graphs is only evaluated with GPU-aware communication.

\section{Performance Evaluation}\label{sec:perf}

In this section, we evaluate the performance and scalability of our approach that
incorporates computation-communication overlap with GPU-aware communication.
We also explore the performance
impact of kernel fusion and CUDA Graphs in strong scaling.

\subsection{Experimental Setup}\label{sec:expr_setup}


We use the Summit supercomputer at Oak Ridge National Laboratory for
conducting our experiments. Summit contains 4,608 nodes each with
two IBM POWER9 CPUs and six NVIDIA Tesla V100 GPUs. Each CPU has
22 physical cores with support for up to four-way simultaneous multithreading (SMT),
contained in a NUMA domain with \mbox{256 GB} of DDR4 memory, totaling \mbox{512 GB} of
host memory. Each GPU has \mbox{16 GB} of HBM2 memory, with an aggregate GPU memory
of 96 GB per node. Summit compute nodes are connected in a non-blocking
fat tree topology with dual-rail EDR Infiniband, which has an injection bandwidth of \mbox{23 GB/s}.
The \mbox{Bridges-2} supercomputer at Pittsburgh Supercomputing Center and Expanse at
San Diego Supercomputer Center have also been used to test and debug GPU acceleration in Charm++.

The performances of the MPI versions of Jacobi3D are obtained using
the default MPI and CUDA environments on Summit: IBM Spectrum MPI 10.4.0.3 and CUDA 11.0.3.
The Charm++ versions of Jacobi3D use the yet-to-be-released Channel API, with UCX 1.11.1 and CUDA 11.4.2.
The more recent version of CUDA used with Charm++ is not compatible with IBM Spectrum MPI,
which is why an older version of CUDA is used for the MPI experiments.
In our tests, we have not observed any noticeable difference
in performance between the two CUDA versions.

As is the norm with GPU-accelerated MPI applications, each MPI process is mapped to one CPU core and one GPU, and is
responsible for a cuboid block of the global simulation grid. For example, when Jacobi3D
is run on a single node (six MPI processes and GPUs), the global grid is divided into six equal-sized blocks;
the grid is decomposed in a way that minimizes the aggregate surface area, which is tied to communication volume.
The Charm++ experiments are also carried out using one CPU core and one GPU per process in non-SMP mode,
but with an additional parameter, \textit{Overdecomposition Factor (ODF)},
which determines the number of chares per PE and GPU. With an ODF of one, the decomposition
of a Charm++ program is equivalent to MPI, where one chare object is mapped to each PE.
A higher ODF creates more chares each with finer granularity,
providing more opportunities for computation-communication overlap and runtime adaptivity,
albeit with increased fine-grained overheads.
We experiment with varying ODFs from one to 16, increased by a factor of two,
to observe the impact of overdecomposition on performance. 

For the following scalability experiments, we compare the performance of four
different versions of Jacobi3D: MPI with host-staging communication (\mbox{MPI-H}),
CUDA-aware MPI (\mbox{MPI-D}),
Charm++ with host-staging communication (\mbox{Charm-H}) and Charm++ with GPU-aware
communication using Channel API (\mbox{Charm-D}).
The Charm++ versions of Jacobi3D are run with different ODFs and the one with
the best performance is chosen as the representative for each point in scaling.
Jacobi3D is run for 10 warm-up iterations and then timed for 100 iterations.
Each experiment is repeated three times and averaged to obtain accurate performance results.

\begin{figure}[t]
	\centering
	\begin{subfigure}{\linewidth}
		\centering
		\includegraphics[width=0.8\linewidth]{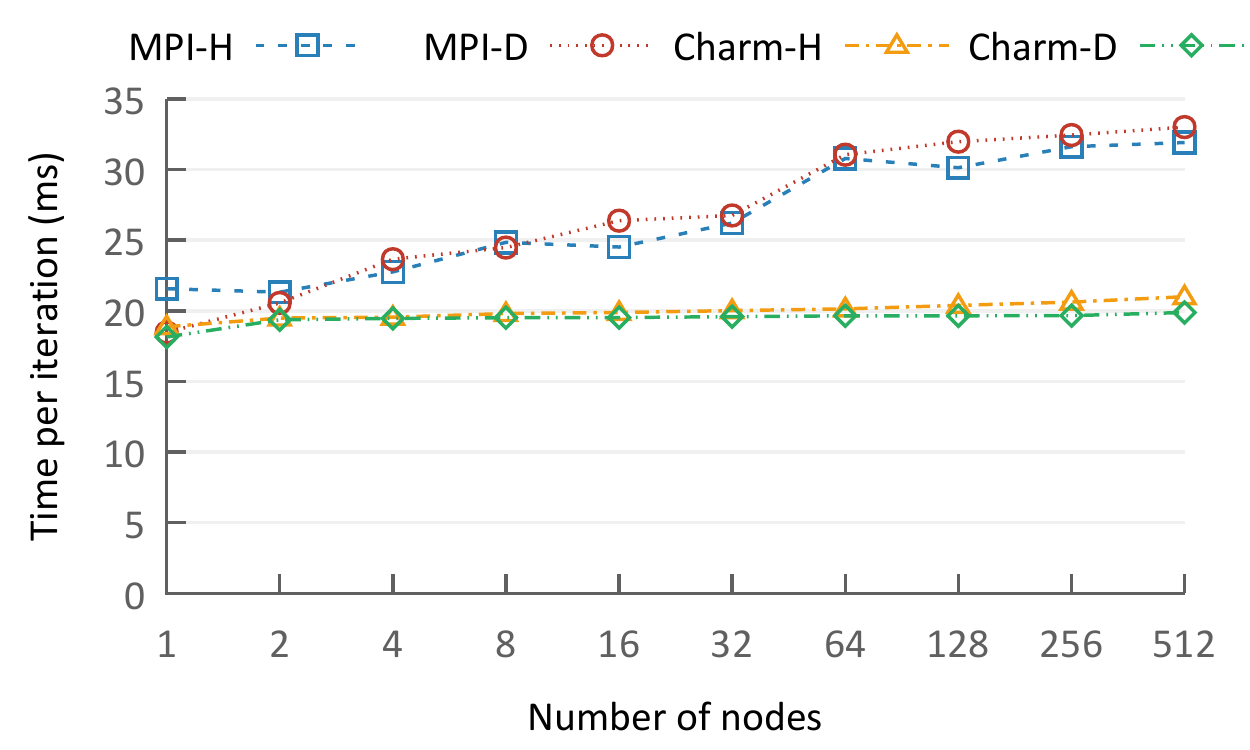}
		\caption{Weak scaling, with block size of 1536 $\times$ 1536 $\times$ 1536 per node}
		\label{fig:jacobi3d_big_weak}
	\end{subfigure}
	\hfil
	\vspace{10pt}
	\begin{subfigure}{\linewidth}
		\centering
		\includegraphics[width=0.8\linewidth]{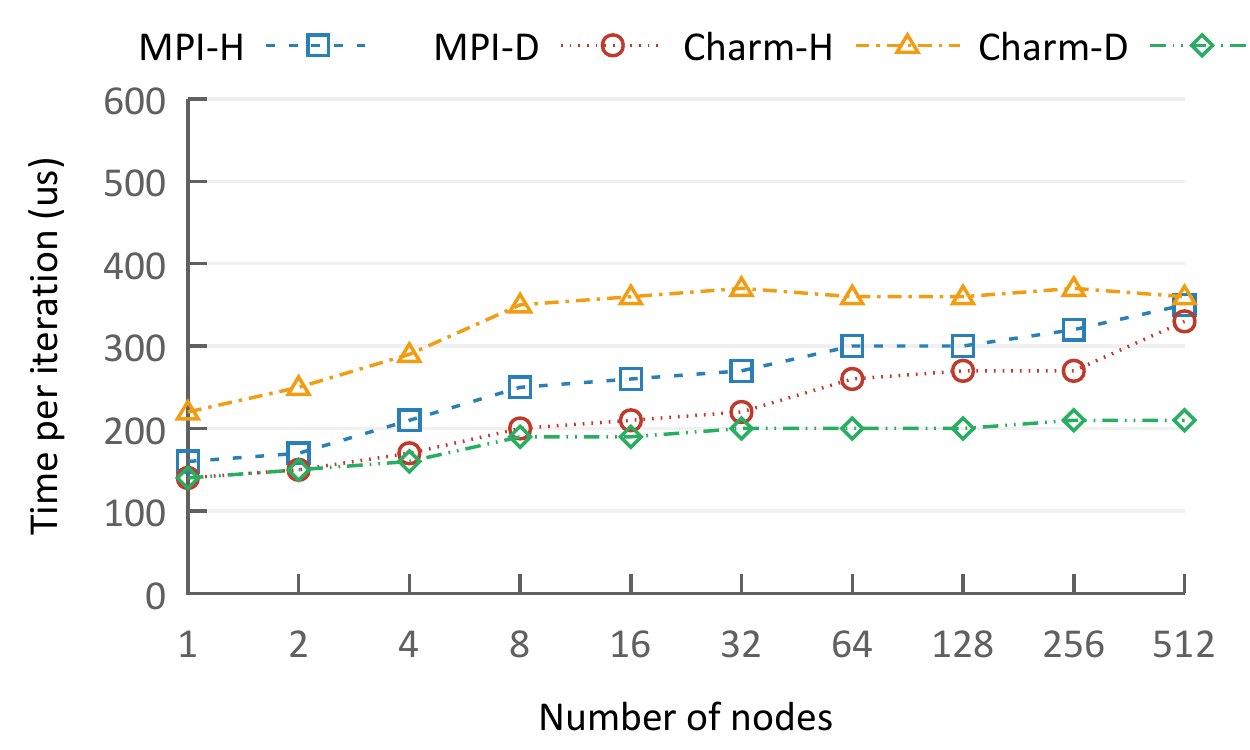}
		\caption{Weak scaling, with block size of 192 $\times$ 192 $\times$ 192 per node}
		\label{fig:jacobi3d_small_weak}
	\end{subfigure}
	\hfil
	\vspace{10pt}
	\begin{subfigure}{\linewidth}
		\centering
		\includegraphics[width=0.8\linewidth]{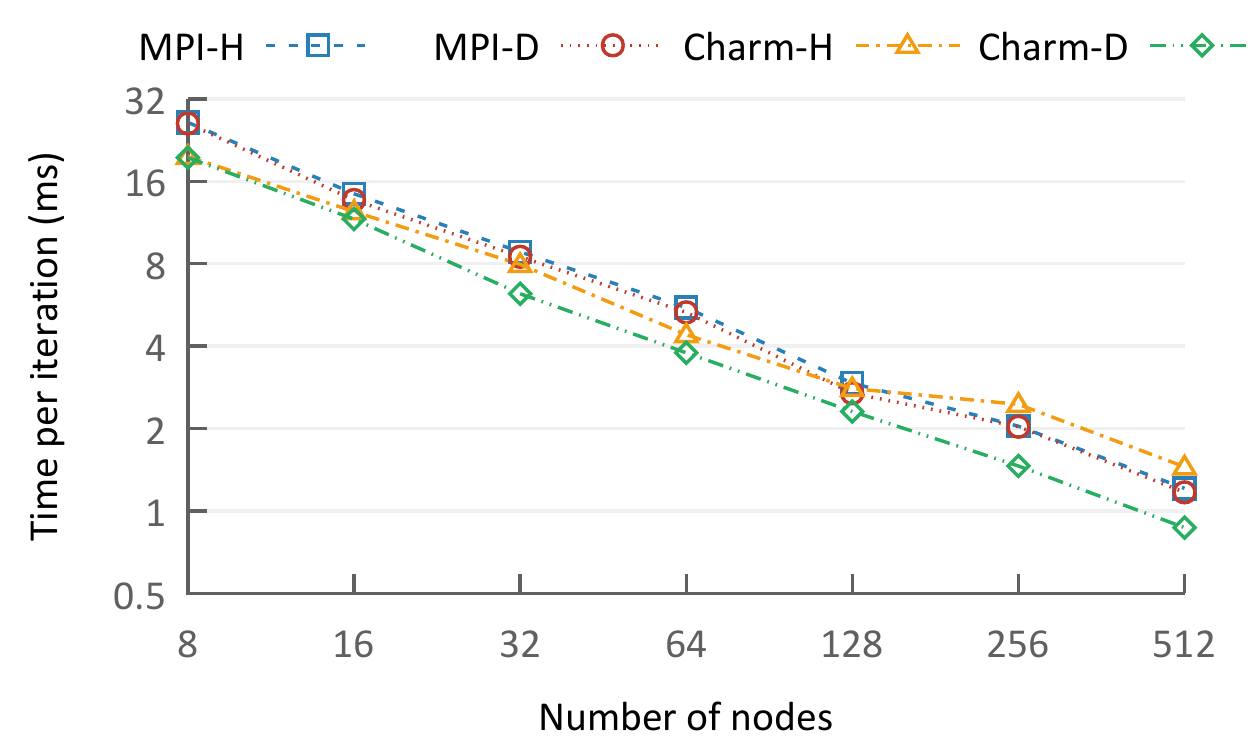}
		\caption{Strong scaling, with global grid size of 3072 $\times$ 3072 $\times$ 3072
			(same global grid size as on eight nodes in Figure~\ref{fig:jacobi3d_big_weak})}
		\label{fig:jacobi3d_big_strong}
	\end{subfigure}
	\caption{Performance comparison of different MPI and Charm++ versions of Jacobi3D.}
	\label{fig:jacobi3d_scalability}
	\vspace{-10pt}
\end{figure}

\subsection{Weak Scaling}\label{sec:weak_scaling}

We evaluate the weak scaling performance of Jacobi3D using two different base problem sizes per node: 1,536 $\times$ 1,536 $\times$ 1,536 and 192 $\times$ 192 $\times$ 192. Each element of the grid is a double precision floating point (eight bytes).
With weak scaling, the size of each dimension is increased successively by a factor of two, allowing the data size per GPU to remain approximately the same.
When decomposed into six GPUs per node, the larger problem size uses roughly \mbox{9 GB} of GPU memory and the smaller problem uses \mbox{18 MB}, most of which is for storing two separate copies of the block data from the previous and current iterations.
The size of messages being exchanged in the halo exchange phase also differs greatly, with up to \mbox{9 MB} and \mbox{96 KB}, respectively.

Figure~\ref{fig:jacobi3d_big_weak} compares the weak scaling performance of the different implementations of Jacobi3D, with a base problem size of 1,536$^3$.
ODF-4 (four chares per GPU) provides the best performance out of all the tested ODFs in Charm-H, whereas ODF-2 performs the best in Charm-D.
These ODFs strike a good balance between computation-communication overlap and overdecomposition overheads; an excessive ODF creates too many fine-grained chares whose overheads can outweigh the benefits from overlap.
Charm-D shows the best performance at a lower ODF than Charm-H, since GPU-aware communication substantially reduces communication overheads
and does not require higher degrees of overdecomposition for more aggressive overlap.
Charm-D outperforms Charm-H only by up to 5\% since the automatic computation-communication overlap employed in Charm-H is able to hide most of the communication overheads.
Nevertheless, combining automatic overlap and GPU-aware communication provides a performance improvement of 61\% on 512 nodes, compared to the performance without overdecomposition and with host-staging communication.

An interesting observation in Figure~\ref{fig:jacobi3d_big_weak} is that GPU-aware communication in IBM Spectrum MPI (MPI-D) does not improve performance starting from four nodes.
By profiling the runs with NVIDIA Nsight Systems, we find that the large message sizes (up to 9 MB) in the halo exchanges cause a protocol change in the underlying communication framework.
For such large messages, a pipelined host-staging mechanism that splits each message into smaller chunks is used,
rather than GPUDirect~\cite{exampi20-hanford}.
Conversely, this behavior does not appear in UCX-based Charm++ and GPUDirect is always used regardless of the message size.
With Charm++, we observe a more gradual, almost flat incline in execution time compared to MPI, owing to computation-communication overlap providing higher tolerance to increasing communication overheads at scale.

For a smaller base problem size of 192 $\times$ 192 $\times$ 192 (halo size of up to \mbox{96 KB}),
GPU-aware communication provides substantial improvements in performance in both MPI and Charm++
as demonstrated in Figure~\ref{fig:jacobi3d_small_weak}.
However, because of the much smaller task granularity, overheads from the Charm++ runtime system
including scheduling chares, location management, and packing/unpacking messages become more pronounced.
Moreover, overdecomposition only degrades performance, as the potential benefits from overlap pale
in comparison to the overheads of finer decomposition; \mbox{ODF-1} (no overdecomposition) performs
the best in both \mbox{Charm-H} and \mbox{Charm-D}.
The performance of CUDA-aware Spectrum MPI (\mbox{MPI-D}) becomes unstable on 64 or more nodes,
with the time per iteration varying between 300 us and 800 us from run to run.
There seems to be a problem with the MPI library as we have been able to reproduce this issue multiple times.

\subsection{Strong Scaling}\label{sec:strong_scaling}

For strong scaling, we experiment with a fixed global grid of size 3,072 $\times$ 3,072 $\times$ 3,072.
As we scale out and the number of nodes is doubled, the size of each work unit decreases by a factor of two.
With Charm++, this means that the best overdecomposition factor will likely become smaller,
as the overheads from high degrees of overdecomposition grow.
Figure~\ref{fig:jacobi3d_big_strong} illustrates the strong scaling performance of the different versions of Jacobi3D.
The best ODF of \mbox{Charm-H} remains at four until 16 nodes, after which \mbox{ODF-2} starts to outperform
until 512 nodes, where \mbox{ODF-1} performs the best. For \mbox{Charm-D}, \mbox{ODF-2} provides the best performance
at all scales, demonstrating that the reduction in communication overheads from GPU-aware communication
enables a higher degree of overdecomposition to retain its effectiveness.
On 512 nodes, \mbox{ODF-2} in \mbox{Charm-H} is 13\% slower than \mbox{ODF-1}, whereas \mbox{ODF-2} in \mbox{Charm-D} is 13\% faster than \mbox{ODF-1}.
The performance issue observed with pipelined host-staging communication in MPI with weak scaling
becomes less relevant with strong scaling, as GPUDirect is used instead at larger scales with
the smaller halo messages.
\mbox{Charm-H}, with host-staging communication, outperforms both \mbox{MPI-H} and \mbox{MPI-D} implementations
until 128 nodes thanks to overdecomposition-driven overlap. \mbox{Charm-D}, combining automatic computation-communication
overlap and GPU-aware communication, substantially outperforms all other versions of Jacobi3D and scales out further,
achieving a sub-millisecond average time per iteration on 512 nodes (3,072 GPUs).

We also evaluate the performance impact of kernel fusion and CUDA Graphs, which are techniques that can be used
to counter fine-grained overheads in strong scaling\footnote{These techniques can also be helpful in weak scaling
with a small base problem size, but we focus on their effects on strong scaling in this work.}.
The Charm++ version of Jacobi3D with GPU-aware communication (\mbox{Charm-D} in previous plots)
is used as the baseline for this experiment, with a relatively small simulation grid of
768 $\times$ 768 $\times$ 768 scaled out to 128 nodes.
In this case, overdecomposition does not improve performance; nevertheless, we present results both
without overdecomposition (\mbox{ODF-1}) and with a high degree of overdecomposition (\mbox{ODF-8}),
to consider scenarios where overdecomposition can be used for other adaptive runtime features
such as dynamic load balancing rather than for performance.

\begin{figure}[t]
	\centering
	\begin{subfigure}{\linewidth}
		\centering
		\includegraphics[width=0.8\linewidth]{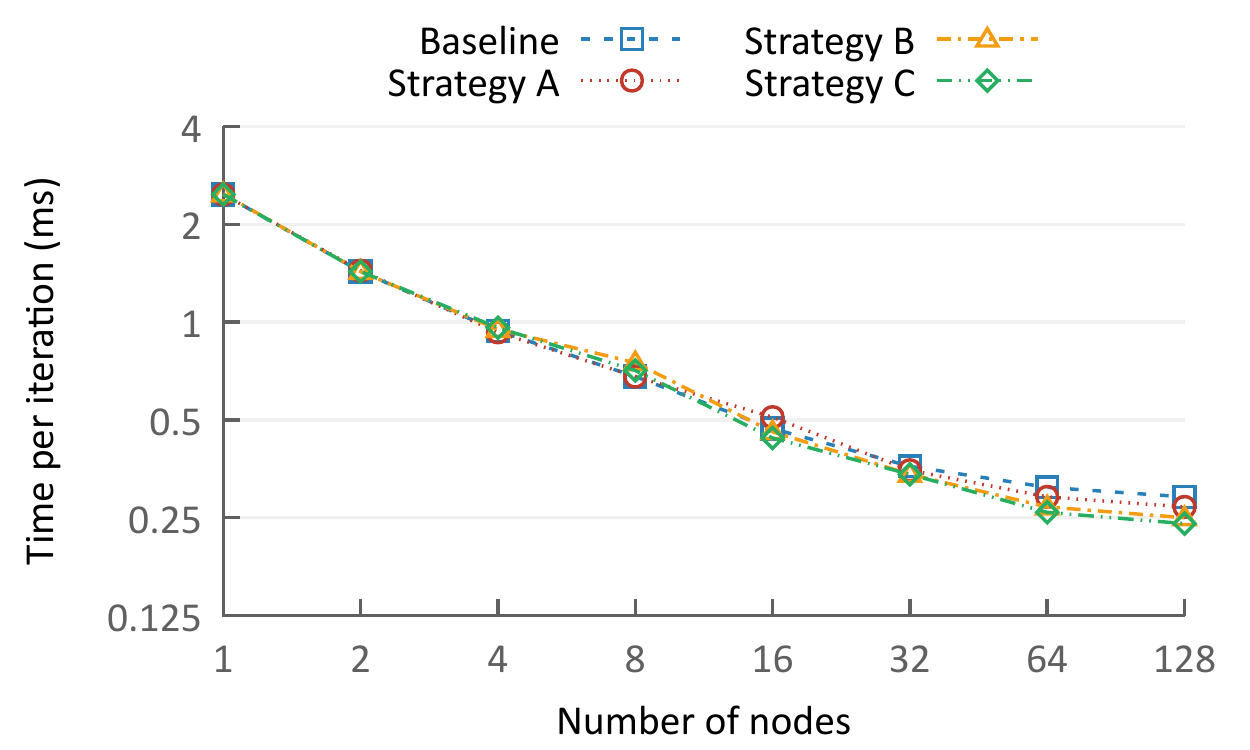}
		\caption{ODF-1}
		\label{fig:kernel_fusion_odf_1}
	\end{subfigure}
	\hfil
	\vspace{10pt}
	\begin{subfigure}{\linewidth}
		\centering
		\includegraphics[width=0.8\linewidth]{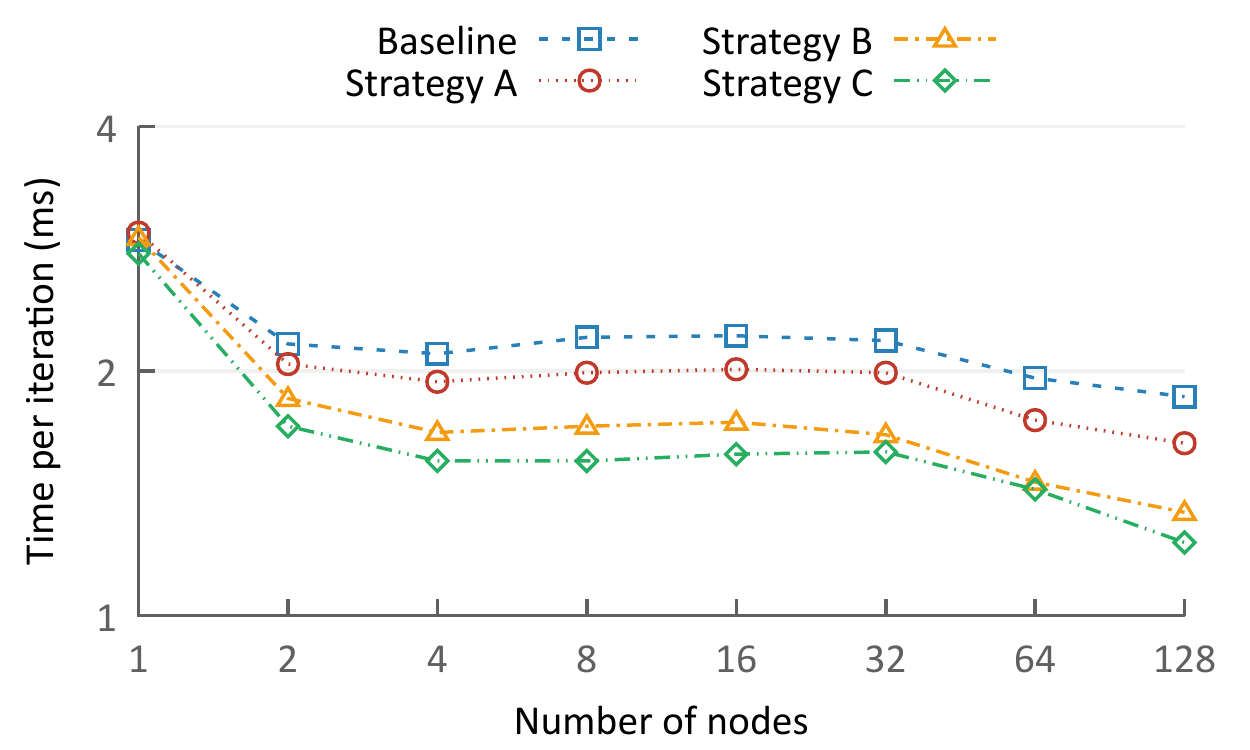}
		\caption{ODF-8}
		\label{fig:kernel_fusion_odf_8}
	\end{subfigure}
	\caption{Impact of kernel fusion on the strong scaling performance of the Charm++ version of Jacobi3D with GPU-aware communication.}
	\label{fig:kernel_fusion}
	\vspace{-10pt}
\end{figure}

\subsubsection{Kernel Fusion}

Figure~\ref{fig:kernel_fusion} illustrates the effectiveness of the kernel fusion strategies described in Section~\ref{sec:design_kernel_fusion} in strong scaling performance. The baseline results do not employ any type of kernel fusion, and fusion strategies from A to C become increasingly aggressive (fusing more types of kernels).
Without overdecomposition (\mbox{ODF-1}), kernel fusion does not noticeably affect performance until 32 nodes.
At larger scales, however, more aggressive fusion strategies (C > B > A) improve performance more than the others; Strategy C improves the average time per iteration by 20\% on 128 nodes.
This demonstrates that kernel fusion is indeed effective at mitigating kernel launch overheads, especially with smaller task granularity at the limits of strong scaling.
Greater performance effects from kernel fusion can be observed with \mbox{ODF-8}, where the already fine-grained work units are further split up with an eight-fold overdecomposition. Fusion strategy C provides up to 51\% increase in the overall performance on 128 nodes.

Although higher degrees of overdecomposition can degrade performance with small problem sizes, they may be needed to enable adaptive runtime features such as load balancing and fault tolerance.
As such, kernel fusion can be a useful technique for reducing kernel launch overheads to improve strong scaling performance especially with overdecomposition.

\begin{figure}[t]
	\centering
	\begin{subfigure}{\linewidth}
		\centering
		\includegraphics[width=0.8\linewidth]{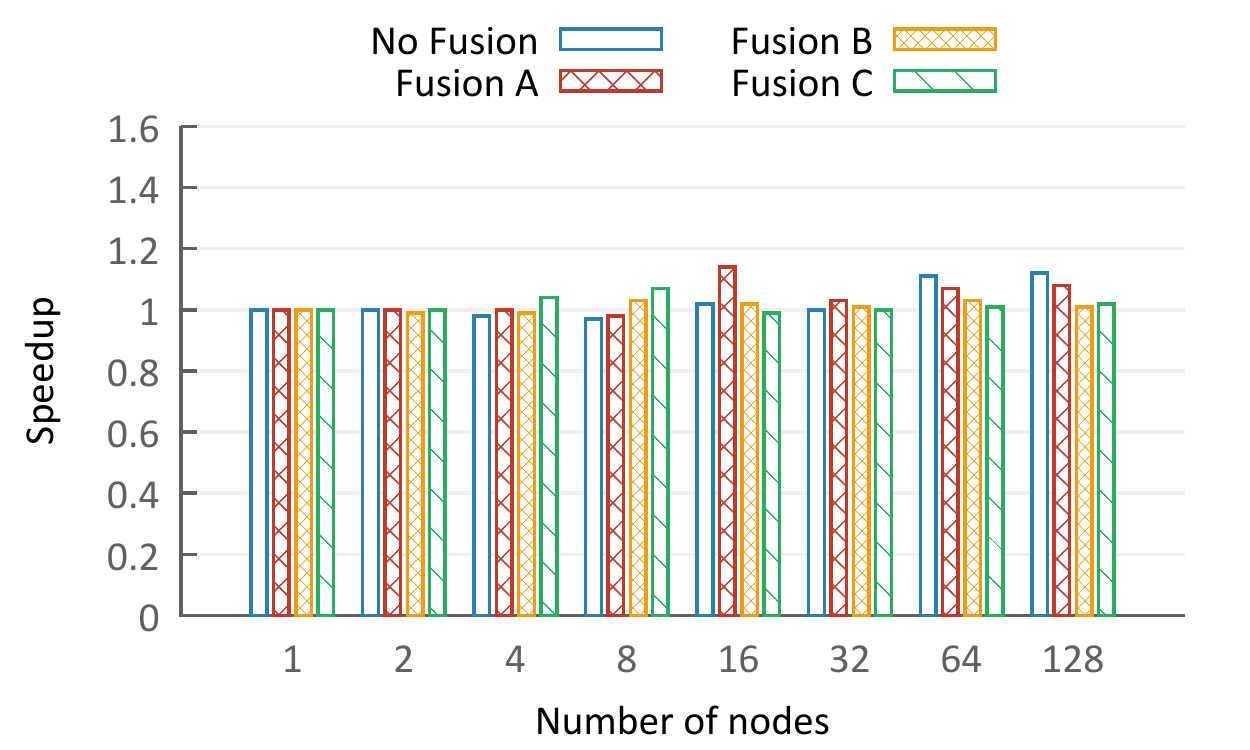}
		\caption{ODF-1}
		\label{fig:cuda_graphs_odf_1}
	\end{subfigure}
	\hfil
	\vspace{10pt}
	\begin{subfigure}{\linewidth}
		\centering
		\includegraphics[width=0.8\linewidth]{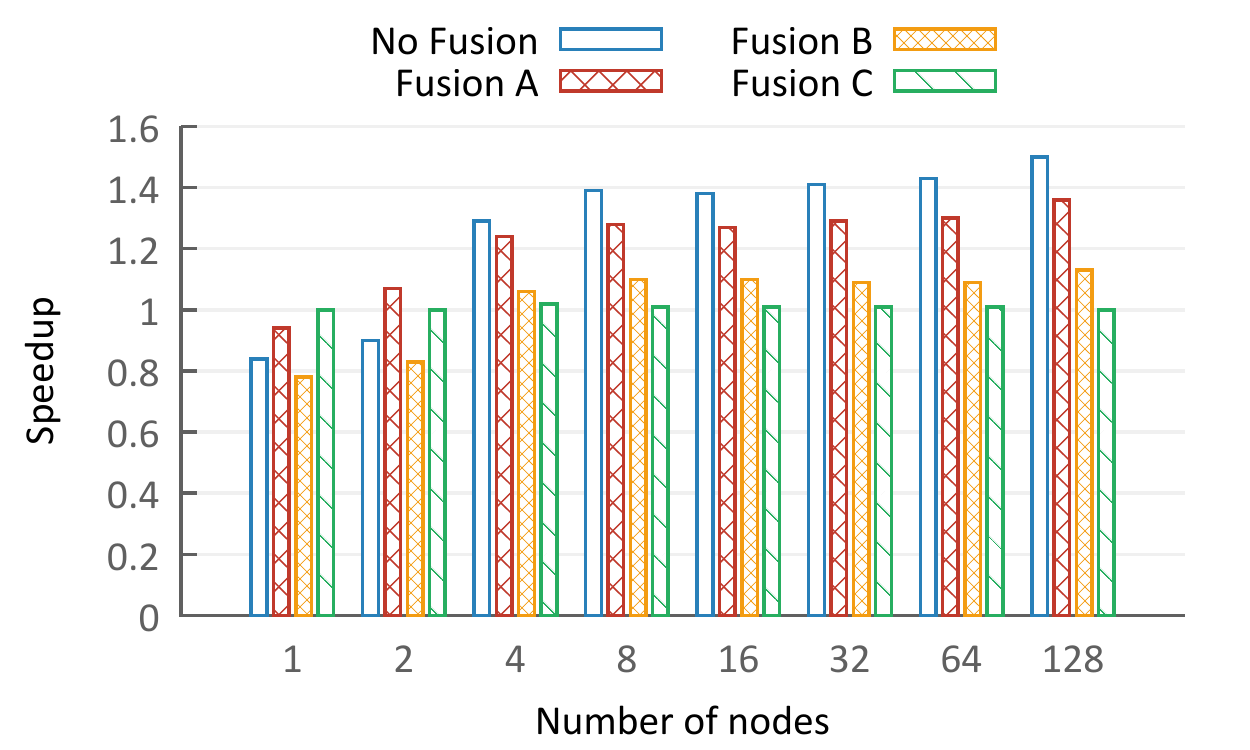}
		\caption{ODF-8}
		\label{fig:cuda_graphs_odf_8}
	\end{subfigure}
	\caption{Speedup from using CUDA Graphs in addition to kernel fusion with GPU-aware Charm++ Jacobi3D.}
	\label{fig:cuda_graphs}
	\vspace{-10pt}
\end{figure}

\subsubsection{CUDA Graphs}

Figure~\ref{fig:cuda_graphs} shows the obtained speedup from using CUDA Graphs, with and without kernel fusion.
Without overdecomposition (\mbox{ODF-1}), CUDA Graphs has little impact on the overall performance, with small improvements
at larger scales and less aggressive fusion strategies. Such moderate performance improvement when compared to
other studies~\cite{pytorch_cuda_graphs} stems from the low CPU utilization in Jacobi3D, where
the CPU resources are mostly used only by the Charm++ runtime system. With bulk of the computation offloaded to the GPU,
CPUs largely sit idle waiting for GPU work to complete, aside from scheduling chares for execution and managing communication.
This causes the reduction in aggregate kernel launch latency from the use of CUDA Graphs to have less impact on the performance of Jacobi3D, when compared to workloads such as deep learning in PyTorch~\cite{pytorch_cuda_graphs} that heavily utilize CPU resources in addition to GPUs.

However, performance improvements are more apparent with ODF-8, where we obtain a speedup of 1.5x on 128 nodes without kernel fusion.
This is because CPU utilization rises substantially in accordance with the increase in overdecomposition factor.
More fine-grained tasks are created, resulting in more kernel launches and GPU operations that utilize the host CPU.
Conversely, the performance impact of CUDA Graphs diminishes as a more aggressive kernel fusion strategy is used, even with ODF-8.
With a higher degree of kernel fusion, the total number of kernels decreases, leaving less room for improvement in the aggregate kernel launch latency.
In summary, CUDA Graphs has the potential to provide substantial performance improvements especially for workloads with high CPU utilization and when there are a sufficient number of kernel launches to optimize.

\section{Related Work}\label{sec:related}


Task-based programming models such as Legion~\cite{legion} and HPX~\cite{pgas14-hpx}
facilitate automatic computation-communication overlap by extracting parallelism
at the level of the runtime system. Castillo et al.~\cite{ics19-overlap} discusses the disparity between
asynchronous task-based programming models and the underlying messaging layer (MPI)
that limits achievable overlap.
A study by Danalis et al.~\cite{sc05-trans}
applies transformations to the application code to expose more opportunities for overlap.
As for GPU-aware communication, many works have discussed the necessary implementations and
improvements in performance~\cite{jcs11-mvapich2-gpu, ipdpsw12-cuda_ipc, icpp13-potluri}.
This work distinguishes itself from others by illustrating the gains in performance and scalability
from combining GPU-aware communication with automatic computation-communication overlap, enabled with overdecomposition.

\section{Conclusion}\label{sec:conclusion}

In this work, we explored how automatic computation-communication overlap from overdecomposition and asynchronous execution
can be used together with GPU-aware communication to improve performance and scalability on modern GPU-accelerated systems.
Using implementations in MPI and Charm++ of a scientific proxy application, Jacobi3D, we evaluated the impact
of our approach on both weak and strong scaling performance with various problem sizes. We observed that the Charm++
version of Jacobi3D with overdecomposition-driven overlap and GPU-aware communication is able to achieve the best
performance with strong scaling, achieving a sub-millisecond time per iteration on 512 nodes of the Summit supercomputer.
With weak scaling, however, we see that the performance impact of combining overdecomposition and GPU-aware communication
varies depending on the problem size.

In addition to demonstrating the importance of minimizing host-device synchronizations
and increasing concurrency in GPU operations, we evaluated the usage of kernel fusion and CUDA Graphs to mitigate
fine-grained execution in strong scaling scenarios. With the most aggressive kernel fusion strategy, we achieved
up to 20\% improvement in overall performance with \mbox{ODF-1} and 51\% with \mbox{ODF-8}.
CUDA Graphs enabled performance improvements of up to 50\% when used without kernel fusion,
demonstrating its effectiveness for workloads with high CPU utilization and a large number of kernel launches.

\section*{Acknowledgment}

\newcommand\blfootnote[1]{%
	\begingroup
	\renewcommand\thefootnote{}\footnote{#1}%
	\addtocounter{footnote}{-1}%
	\endgroup
}

This work was performed under the auspices of the U.S. Department
of Energy (DOE) by Lawrence Livermore National Laboratory
under Contract DE-AC52-07NA27344 (LLNL-CONF-832823).

This
research was supported by the Exascale Computing Project (17-SC-20-SC),
a collaborative effort of the U.S. DOE Office of Science and
the National Nuclear Security Administration.

This research used resources of the Oak Ridge Leadership Computing
Facility at the Oak Ridge National Laboratory, which is
supported by the Office of Science of the U.S. DOE under Contract
No. DE-AC05-00OR22725.

This work used the Extreme Science and Engineering Discovery Environment (XSEDE), which is supported by National Science Foundation grant number ACI-1548562. XSEDE resources include Bridges-2 at Pittsburgh Supercomputing Center (PSC) and Expanse at San Diego Supercomputer Center (SDSC), used through allocation TG-ASC050039N.


\bibliographystyle{IEEEtran}
\bibliography{IEEEabrv,ref}

\end{document}